\documentclass[twocolumn,showpacs,amsmath,amssymb,aps,pra,floatfix,nofootinbib,10pt]{revtex4-2}
\usepackage[T2A]{fontenc}
\usepackage[utf8]{inputenc}
\usepackage[english
]{babel}
\usepackage[T1]{fontenc}
\usepackage{mathrsfs}
\usepackage{hyperref}
\usepackage{multirow}
\usepackage{bm,epsfig}
\usepackage{graphicx}
\usepackage{subeqnarray}
\usepackage{bbold}
\usepackage{upgreek}
\hypersetup{colorlinks=true, urlcolor=blue, citecolor=blue, filecolor=blue, linkcolor=blue}
\usepackage{dcolumn}
\usepackage{amsmath}
\usepackage{color}
\usepackage{siunitx}
\usepackage{physics}

\usepackage{xcolor}

\renewcommand{\Re}{\operatorname{Re}}
\renewcommand{\Im}{\operatorname{Im}}

\def\beq#1{\begin{equation}\label{#1}}
\def\eeq{\end{equation}}

\begin{document}

\title{Fano resonance in XUV generated by helium with few-cycle intense laser pulses and its classical analogy} 

\author{
 S.~A.~Bondarenko$^{1,2}$ and V.~V.~Strelkov$^{1,3,\ast}$
}
\affiliation{
\mbox{$^{1}$P.N. Lebedev Physical Institute of the Russian Academy of Sciences, 53 Leninskiy Prospekt,  Moscow 119991, Russia} \\
\mbox{$^{2}$National Research Nuclear University MEPhI, 31 Kashirskoe Highway, Moscow 115409, Russia} \\
\mbox{$^{3}$A. V. Gaponov-Grekhov Institute of Applied Physics of the Russian Academy of Sciences,} \\
{46 Ulyanov street, Nizhny Novgorod 603950,	Russia} \\
$^{\ast}$strelkov.v@gmail.com
}
\begin{abstract}
We integrate numerically the Schr\"odinger equation for the model helium atom irradiated by intense few-cycle laser pulse and find the emitted XUV spectra. They demonstrate resonant peaks at the frequencies of transitions from the doubly-excited autoionizing states (AISs) to the ground state. We study the properties of these peaks depending on the laser pulse duration and find that the decay of the AISs due to photoionization by the laser field affects them. Moreover, we consider the classical system of two coupled oscillators and find that both the quantum (the atom with AIS in the field) and the classical (the coupled oscillators with friction) systems demonstrate Fano-like resonant peak described by an essentially complex asymmetry parameter. We find a remarkable similarity in the behavior of these systems and conclude that the classical system of coupled oscillators with friction is an analogy of the AIS having an extra decay channel in addition to the autoionization one.  
\end{abstract}
\maketitle

\section*{Introduction}

High-order harmonic generation (HHG) of intense laser field is a promising tool to obtain coherent extreme ultraviolet radiation (XUV) in femtosecond or attosecond time domain~~\cite{Villeneuve2018, Ryabikin2023}.  However, the typical efficiencies of the HHG process remain below the level required for many applications. One of the ways to increase the efficiency is using resonances of the generating particles. A very pronounced enhancement of the resonant generation was observed in HHG in plasma plume~\cite{Ganeev2013, GaneevPRA2013, Haessler_2013, Singh2021, Singh2023} and in xenon~\cite{Shiner2011}. Moreover, resonant features were also observed in XUV generated in argon~\cite{Rothhardt2014} and helium~\cite{DOG}. In these papers they observed enhanced generation of XUV with frequency close to the frequency of a transition from an autoionizing state (AIS) to the ground state of the generating atom or ion.  

Several theoretical approaches were suggested to describe this phenomenon~\cite{Milosevic2007, Kheifets2008, Morishita2008, Frolov2009, Strelkov2010, Strelkov2014, Wahyutama2019, Bray2020, Romanov2024}. In particular, paper~\cite{Strelkov2014} generalizes the non-resonant HHG theory~\cite{Lewenstein} to the case when the generating particle has an AIS. It was shown that the XUV spectrum emitted by such a system is a product of the non-resonant spectrum $d_{nr}(\omega)$ and the resonant factor $F$:
\begin{equation}
        d (\omega) = d_{nr}(\omega)  F,
        \label{eq:dipole_moment}
\end{equation}    
where 
\begin{equation}
        F(\epsilon) = 1+Q\frac{1}{\epsilon + i },
        \label{eq:fano_line}
\end{equation}
$\epsilon=\frac{\Delta \omega}{\Gamma/2}$, $\Delta \omega= \omega - \Omega$ is a detuning from the resonance, $\Omega$ is the resonant frequency, $\Gamma$ is the inverse lifetime of the AIS, $Q$ is a complex parameter. 

Eq.~(\ref{eq:fano_line}) describes an asymmetric peak similar to the one found by Fano~\cite{Fano1961}:
\begin{equation}
        T(\epsilon) = \frac{(\epsilon+q)^2}{1+\epsilon^2},
        \label{eq:fano_line1}
\end{equation}
where $q$ is a real Fano parameter. 
For certain $Q$, namely for $Q=q-i$, where $q$ is real, expressions~(\ref{eq:fano_line}) and~(\ref{eq:fano_line1}) are related as $|F|^2=T$. Thus, for this value of $Q$ expression (\ref{eq:fano_line}) describes a 'usual' Fano peak with the (real) asymmetry parameter $q$. The resonant factor is zero at point $\epsilon=-q$ and there is a phase jump at this point. For other $Q$ the resonant factor is non-zero everywhere, and there is no phase jump. 

Note that considering a process with spectrum given by Eq.~(\ref{eq:fano_line1}), the authors of Ref.~\cite{Pfeifer} have shown that in the temporal domain this process is a sum of the Dirac delta function and an exponentially decaying term with the phase $2 \arg(q-i)$. Accordingly, Eq.~(\ref{eq:fano_line}) can be understood as a spectrum of the process which is a sum of the delta function and an exponentially decaying term whose amplitude and phase are defined by the complex parameter $Q$. 

Authors of~\cite{Joe2006} suggested a simple classical system demonstrating a Fano resonance and discussed its origin. The system consists of two oscillators coupled by a spring; one of the oscillators is driven by an external periodic force. In~\cite{Joe2006} they show that the dependence of the amplitude of this oscillator on the driving frequency has two maxima. One of them is symmetric and is located near the natural frequency of this oscillator. The other one is asymmetric and takes place near the natural frequency of the second oscillator. Surprisingly, the width of the latter is non-zero even in the absence of friction in the second oscillator. It this case the width is defined by the coupling between the oscillators. 

In this paper we numerically simulate the XUV spectrum emitted by a model helium atom in a short (few-cycle) laser pulse. The model atom has autoionizing states and the emitted spectra demonstrate an asymmetric resonant maximum at the frequency of the transition from the ground state to AIS. We analytically study the properties of the maximum as a function of the laser pulse duration and compare it with the experimental data. Moreover, we study the behavior of the asymmetric maximum in the system of coupled oscillators under non-zero friction in the second oscillator and find a remarkable similarity of behaviors of the quantum (atom with AIS in the field) and classical (coupled oscillators) systems.

\section*{Numerical methods}
\subsection*{XUV spectrum}

Two-electron atoms are the simplest systems that have autoionizing states. In particular, one-dimensional helium model atom has such states~\cite{Zhao2012}.  We use this model to study the features caused by the autoionization states in an XUV spectrum generated by a short laser pulse. The time-dependent Schr\"odinger equation (TDSE) for 1D helium in an external field~$E(t)$ is written as follows (atomic units are used throughout the paper: $e=m=\hbar=1$):

{\small{
\begin{multline}
     i \frac{\partial}{\partial t}\psi(x, y, t) = \hat{H} \psi(x, y, t), 
     \\ \hat{H} = -\frac{1}{2}\frac{\partial^2}{\partial x^2}-\frac{1}{2}\frac{\partial^2}{\partial y^2}+V(x, y)-i W(x,y)+(x+y)E(t),
    \label{eq:schrodinger} 
\end{multline} } }

where $x$, $y$ are the electrons' coordinates, $V(x, y)$ is an atomic potential, $W(x, y)$ is an absorbing potential, which is non-zero only near the boundaries of the numerical box. Due to this potential the wave function near the boundaries is absorbed and thus it is not reflected back. The atomic potential is: 
\begin{equation}
    V(x,y)=-\frac{2}{\sqrt{x^2+a^2}}-\frac{2}{\sqrt{y^2+a^2}}+\frac{1}{\sqrt{(x-y)^2+b^2}},
\end{equation}

where $a=1/\sqrt{2}$, $b=1/\sqrt{3}$ are constants used in~\cite{Strelkov_dark_2023} (see also~\cite{Javanainen,Zhao2012})  to reproduce the first and second ionization potentials of the three-dimensional helium.

The Schr\"odinger equation~(\ref{eq:schrodinger}) is solved numerically with the method described in~\cite{Strelkov_2006}. The TDSE is solved in the region $200 \times 200$ a.u., the spatial step is $dx=dy=0.2~a.u.$, the absorbing boundary layer (where $W \ne 0$) is located at 85-100 a.u. from the origin.

The laser field is given as:
\begin{equation}
E=-\frac{1}{c}\frac{\partial A}{\partial t}
\label{field}
\end{equation}
where
\begin{equation}
A=E_0 \frac{c}{\omega_{las}} f(t) \sin\left(\omega_{las} t+ \varphi_{CEP}\right),
\label{vekt_pot}
\end{equation}
$\varphi_{CEP}$ is a carrier-envelope phase and $f(t)$ is a slowly-varying pulse envelope. It is given by:
\begin{equation}
    f(t)=
 \begin{cases}
   \cos^2(\frac{\pi}{2}\frac{t}{\tau}) &\text{if $|t|\le \tau$}, 
   \\
   0 &\text{if $|t|>\tau$}.
 \end{cases}
 \label{env}
\end{equation}

Note that for the laser field given by Eqs.~(\ref{field})-(\ref{env}) we have $\int_{-\infty}^{+\infty} E dt=0$ as it should be for an actual laser pulse. 

The TDSE solution is done at the time interval $T$ for $- \tau <t<T-\tau$. The XUV emission takes place during the laser pulse and also after it within the AIS lifetime. We have chosen $T$ long enough to completely describe this process, namely $T=48 T_{las}$ where $T_{las}=\frac{2 \pi}{ \omega_{las}}$ is the laser cycle duration. 

We consider the laser pulse duration $\tau_{FWHM}=\frac{4}{\pi} \tau \arccos{\left(2^{-1/4}\right)} $ from 1.5~fs to 7~fs. The peak laser intensity is $5.6 \times 10^{14}$ W/cm$^2$, the laser wavelength is 800~nm.

\subsection*{Fitting of the resonant line}

Spectrum of the dipole response found via the numerical TDSE solution has resonant features near the frequencies of the transitions from the ground state to the bright AISs (the bright state is the one for which a transition from the ground state is allowed). In this subsection we describe fitting of these features with a Fano-like line~(\ref{eq:fano_line}). 

Let us consider a narrow frequency range near the resonance $\{\Omega - \delta \omega, \Omega + \delta \omega \}$, where $\Omega$ is the resonant frequency, $\delta \omega$ is several units of the resonant width $\Gamma$. Having in mind that we deal with very short laser pulses we suppose that $d_{nr} (\omega)$ in Eq.~(\ref{eq:dipole_moment}) is a continuous spectrum within this frequency range, and the spectral amplitude and phase vary linearly in this range:
\begin{equation}
    d_{nr} (\omega) = (A+B\omega) e^{i C \omega},
\end{equation}
where $A$, $B$ and $C$ are some constants.

Thus from Eq.~(\ref{eq:dipole_moment}) having the numerical spectrum $d_{num} (\omega)$ we reconstruct the resonant factor:
\begin{equation}
    F_{num} (\omega) = d_{num} (\omega) / \left((A+B\omega) e^{i C \omega} \right).
    \label{F_num}
\end{equation}

To find the constants $A$, $B$ and $C$ we take into account that  $\delta \omega$ exceeds $\Gamma$ so from Eq.~(\ref{eq:fano_line}) we have  $\arg(F(\Omega \pm \delta \omega)) \approx 0$ and $|F (\Omega \pm \delta \omega) |^2 \approx 1$. Assuming $\arg(F_{num}(\Omega \pm \delta \omega))=0$ and $|F_{num} (\Omega \pm \delta \omega) |^2=1$ we find constants $A$, $B$ and $C$.

Further we find $Q$ fitting $F_{num} (\omega)$  with Eq.~(\ref{eq:fano_line}) using the gradient descent method. In more details,  for a set of parameters $\Vec{p}=\{\Re(Q),~\Im(Q),~\Gamma,~\Omega \}$ we find the function $F(\omega,\Vec{p} )$ with Eq.~(\ref{eq:fano_line}) and the deviation $\int_{\Omega-\delta \omega}^{\Omega+\delta \omega}  |F_{num} (\omega)- F(\omega,\Vec{p} )|^2 d \omega$. This deviation is minimized using the gradient descent method with $10^5$ different values of the initial parameters $\Vec{p}$, 5000 steps are performed for each set of the initial parameters. 

\section*{Numerical results}
\begin{figure}
    \centering
    \includegraphics[width=0.9\linewidth]{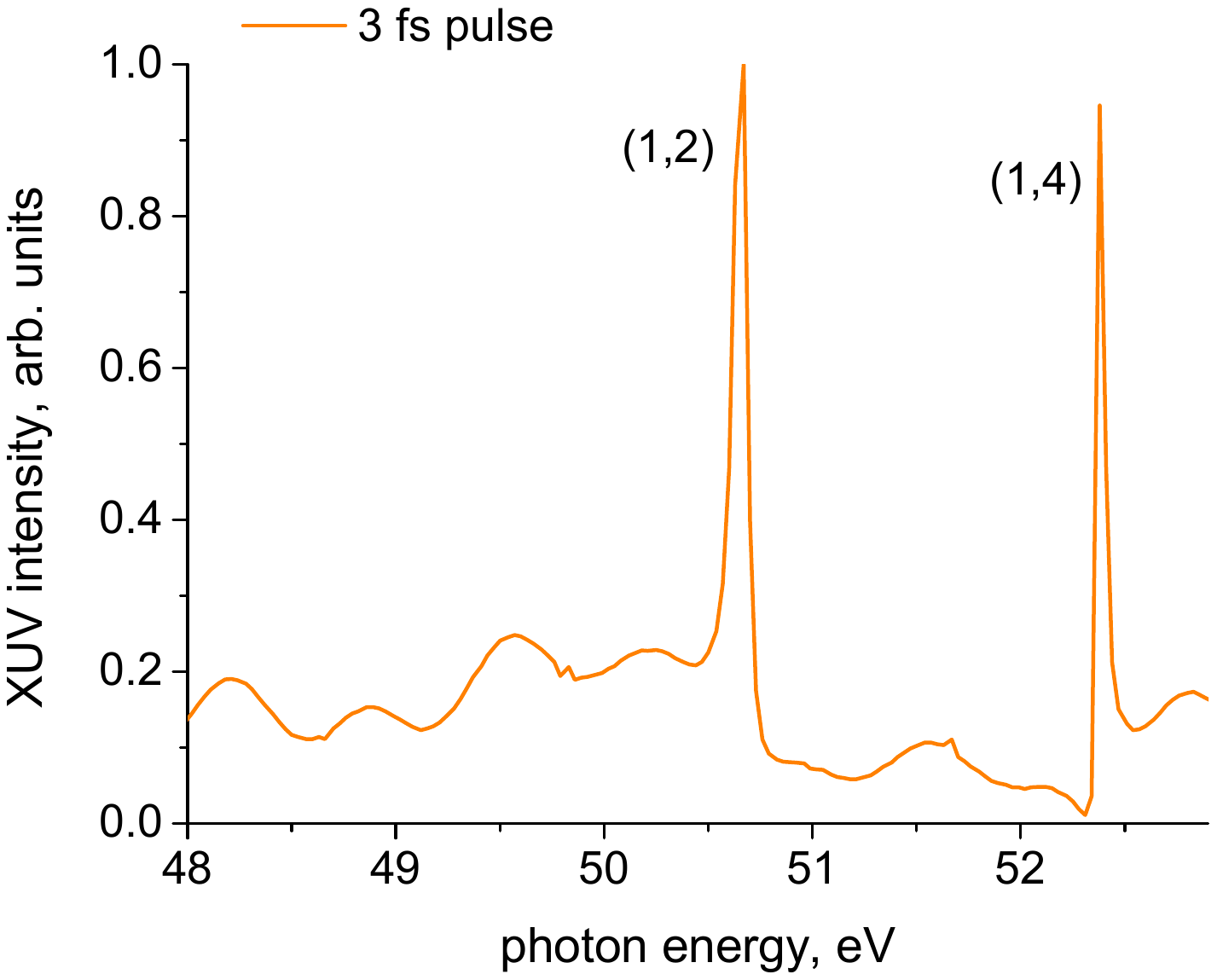}
    \caption{The XUV spectrum near two lowest resonances with the bright AI states. The calculation is done for a cosine-like laser pulse with $\tau_{FWHM}=3$~fs, the central wavelength is 800~nm and the peak intensity is $5 \, \, 10^{14}$W/cm$^2$.  
    }
    \label{fig:spectrum}
\end{figure}

Below we present the results of calculations done for the laser pulses with different durations, the central wavelength of 800~nm, and peak intensity of $5 \, \, 10^{14}$W/cm$^2$. The CEP corresponds to the cosine-like pulse ($\varphi_{CEP}=0$) unless otherwise specified.
Fig.~\ref{fig:spectrum} shows the XUV spectrum in the vicinity of resonances with the two lowest bright AI states. Using the notation of~\cite{Strelkov_dark_2023}, these are (1,2) and (1,4) states. One can clearly see an XUV generation enhancement due to the resonances.

\begin{figure}
    \centering
    \includegraphics[width=\linewidth]{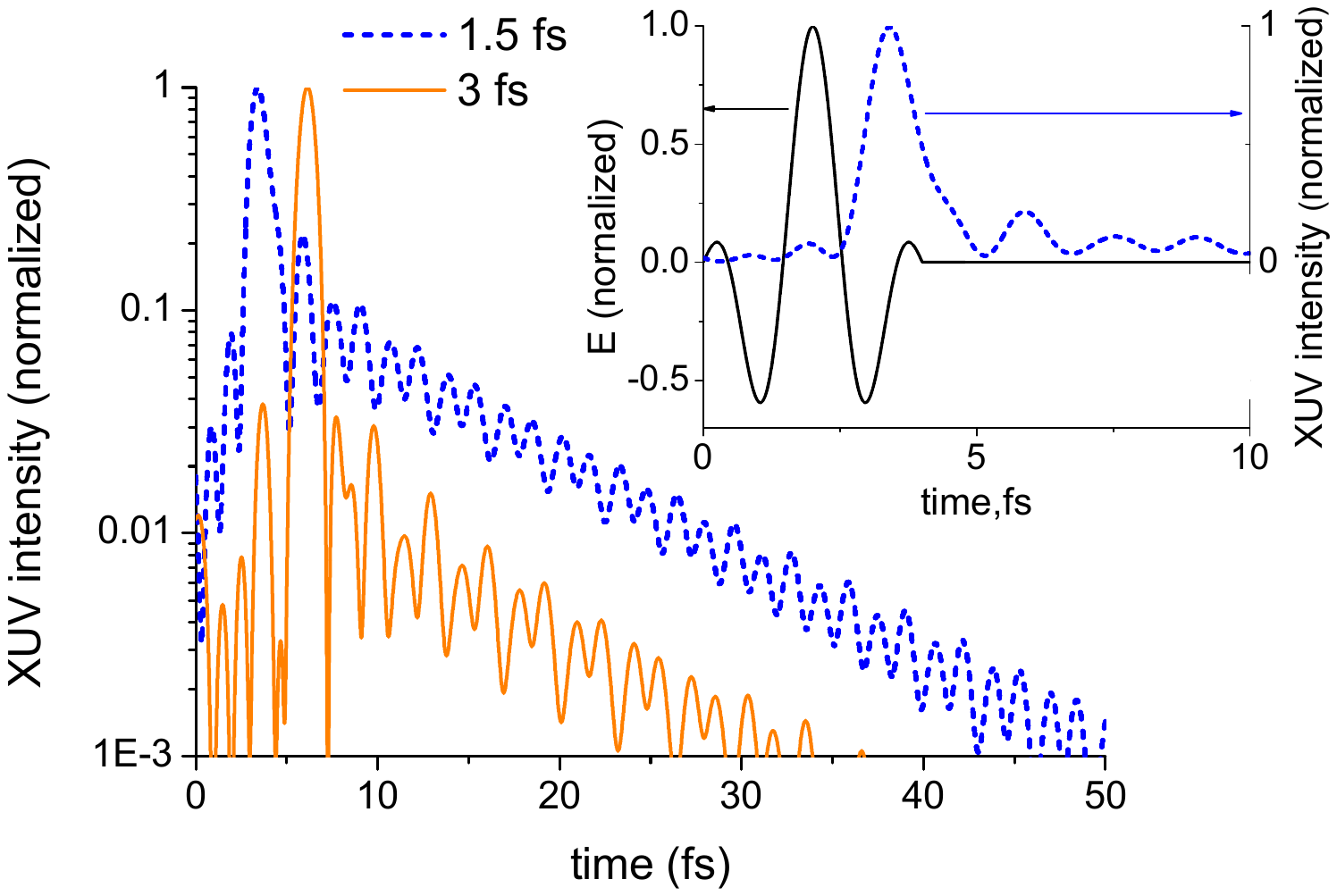}
    \caption{The intensity of the XUV in the frequency range from 48~eV to 52~eV as a function of time for the laser pulse duration 1.5~fs (blue dashed line) and 3~fs (orange solid line). The inset shows the instantaneous laser field strength and the XUV intensity for the laser pulse duration 1.5~fs in the linear scale.  
    }
    \label{fig:temporal_dynamics}
\end{figure}

Fig.~\ref{fig:temporal_dynamics} presents the temporal dynamics of XUV emission enhanced due to resonance with (1,2) state for two different durations of the laser pulse. One can see that this temporal dynamics agrees with the one described in the Introduction. Namely, there is a short flash followed by long exponential decay. The flash corresponds to the XUV emission at the instant of rescattering, and the exponential tail is the emission at the AIS-ground state transition decaying due to the autoionization. Note that these two stages correspond in the spectral domain to the two terms in Eq.~(\ref{eq:fano_line}): the short flash corresponds to the white spectrum ("1" in Eq.~(\ref{eq:fano_line})) and the exponentially decaying emission corresponds to the second term, i.e.the Lorentz line.

\begin{figure}
    \centering
    \includegraphics[width=0.9\linewidth]{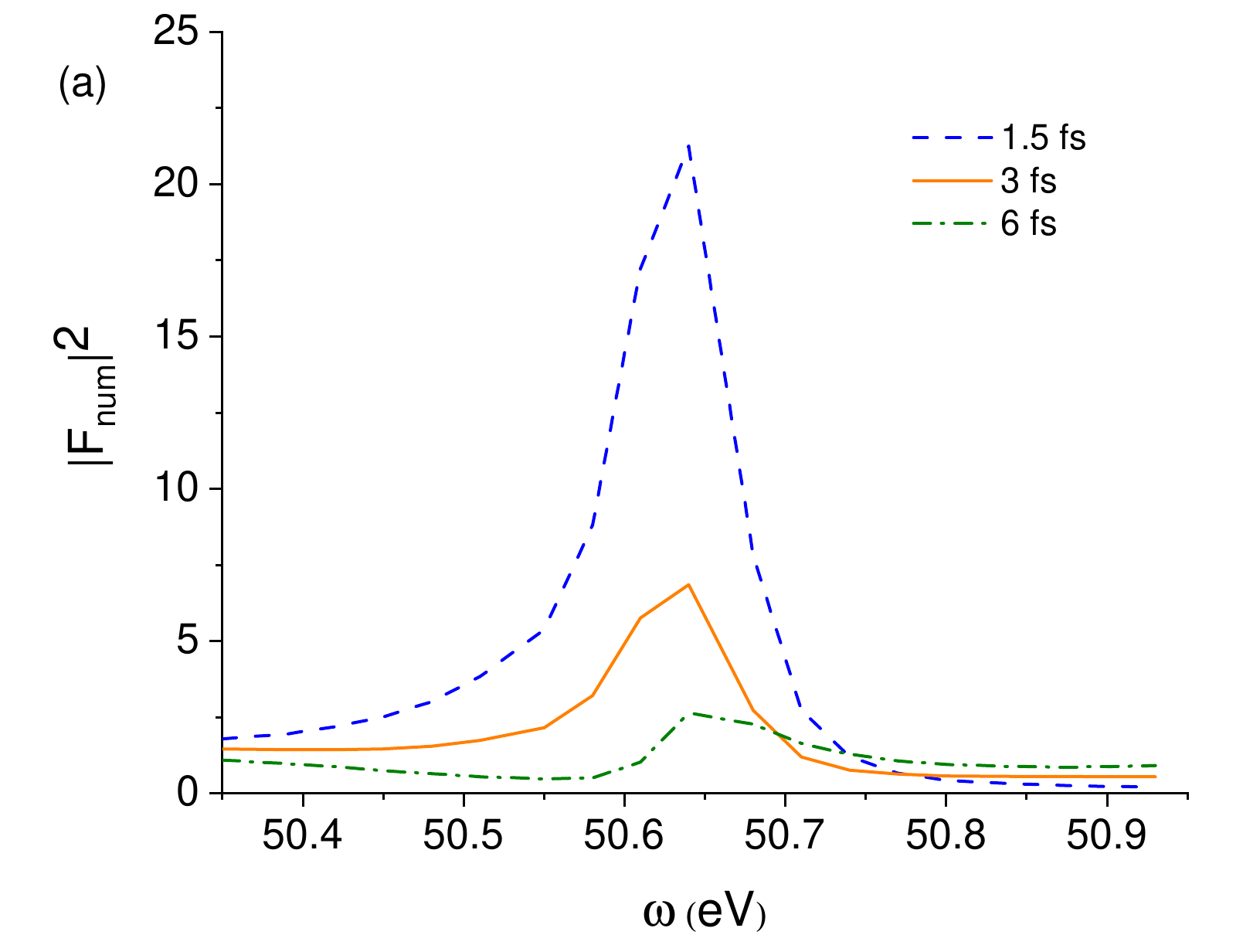}
    \includegraphics[width=0.9\linewidth]{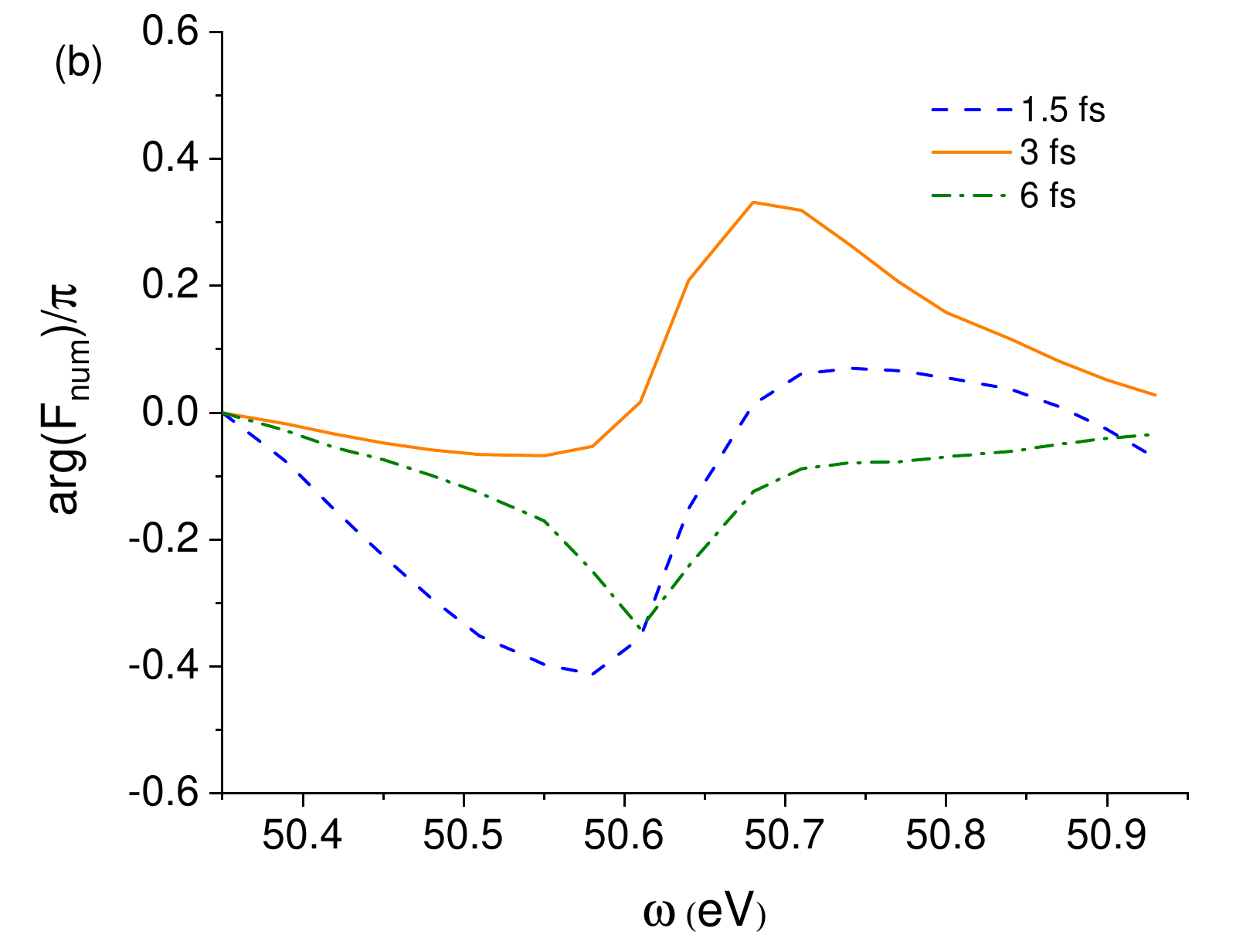}
    \caption{(a) Squared absolute value (a) and phase (b) of the resonant factor found via numerical TDSE solution (see text for more details) for different laser pulse durations shown in the graph.}
    \label{fig:abs_phase_num_3}
\end{figure}

Fig.~\ref{fig:abs_phase_num_3} shows the squared absolute value and the phase of $F_{num} (\omega)$ near the frequency of the transition from the ground state to the lowest bright AIS.  One can see that an increase in the pulse duration leads to an overall decrease of the resonant factor, as well as to a change in the resonant line shape.

\begin{figure}[h]
\includegraphics[width=0.9\linewidth]{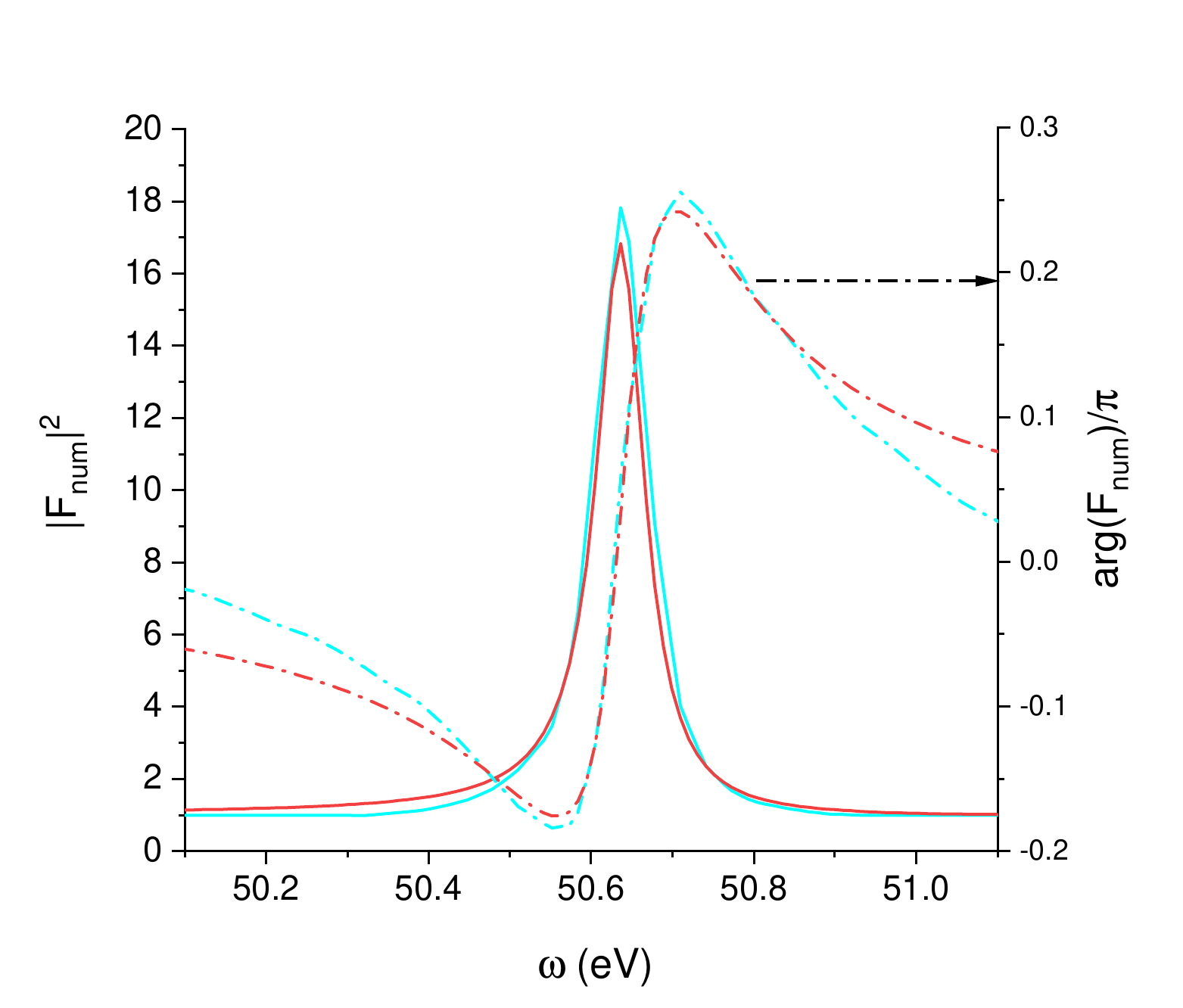}
\caption{The resonant factor $F_{num} (\omega)$ (black lines) and its fitting via Eq.~(\ref{eq:fano_line}) (red lines). The laser pulse duration $\tau_{FWHM}=2$~fs.}
\label{fig:fit_dur_2} 
\end{figure}

Fig.~\ref{fig:fit_dur_2} illustrates the result of the fitting procedure. We show the squared absolute value and the phase of the resonant factor $F_{num} (\omega)$ and its fitting via Eq.~(\ref{eq:fano_line}). The parameters of the fitting curve found via the gradient descent method are $\Re(Q)=-0.53$, $\Im(Q)=3.07$, $\Gamma=0.07$~eV, $\Omega=50.64$~eV.

\begin{figure}[h]
    \centering
    \includegraphics[width=0.9\linewidth]{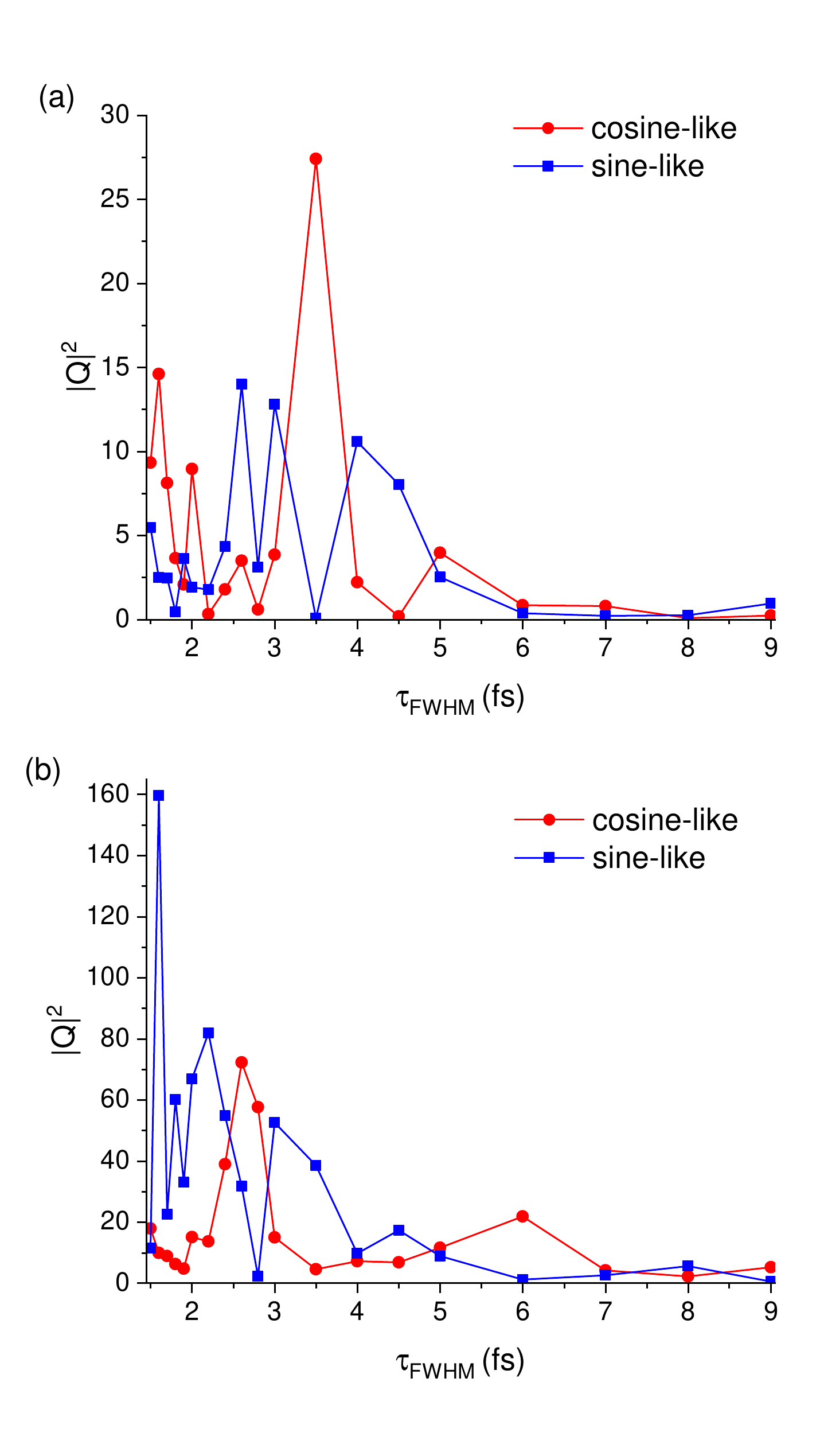}
    \caption{The calculated $|Q|^2$ as a function of the laser pulse duration for cosine-like (red circles) and sine-like (blue squares) pulses. Results for the resonance with (1,2) state (a) and with (1,4) state (b).}
    \label{fig:abs_q}
\end{figure}

The resonant factor $F_{num} (\omega)$ was calculated and fitted via Eq.~(\ref{eq:fano_line}) for different laser pulse durations, for two CEP values ($\varphi_{CEP}=0$ and $\pi/2$), for two resonances, shown in Fig.~\ref{fig:spectrum}. Results are shown in Fig.~\ref{fig:abs_q}. One can see that up to the durations of approximately 3-6~fs the behavior of this value is complicated. This complexity can be attributed to constructive or destructive interference of contributions of different half-cycles of the laser field to the XUV emission. However, for longer laser pulses $|Q|^2$ monotonically decreases with the laser pulse duration for both resonances and both CEPs. This behavior can be attributed to the photoionization of the autoionizing state: when the pulse duration is long, the AIS is populated via rescattering but depopulated at the successive half-cycles due to photoionization; when the laser pulse is very short, this is not the case because the laser field vanishes soon after the rescattering. This feature can be seen in Fig.~\ref{fig:temporal_dynamics}: in the case of 1.5~fs laser pulse the exponential tail is much more intense than in the case of 3~fs pulse.

 The behavior we found for the resonant peaks in the XUV spectrum agrees with the experimental results on HHG in helium using few-cycle laser pulses~\cite{DOG}. In this work they observed the resonant feature in the continuous XUV spectrum at the frequency of the  transition from the ground to AI state. The feature was very pronounced for the shortest laser pulse and became less pronounced with the laser pulse duration increase. 

The high sensitivity of the AIS to the laser field agrees with results of Ref.~\cite{Pfeifer} where they have observed the change of the  resonant line shape in the absorption spectrum of helium using two-orders of magnitude weaker few-cycle laser pulse. Such high sensitivity is natural for the AIS in helium which is a doubly-excited state: in the latter both electrons are far from the nucleus so their motion can be easily affected by the laser field. The onset of strong-field effects on the doubly-excited states in helium was studied very recently in~\cite{Rupprecht2024}. Note that other type of AISs, namely the one originating from the vacancies in inner orbitals, should be less sensitive to the laser field. Accordingly, resonant HHG enhancement due to such AISs is observed using many-cycle laser pulses~\cite{Ganeev2013, GaneevPRA2013, Haessler_2013, Singh2021, Singh2023, Shiner2011, Rothhardt2014}.   

\section*{Coupled oscillators}

Let us consider two classical oscillators coupled by a spring; one of the oscillators (denoted below as the first one) is driven by a periodic force. The equation  of motion for the oscillators are written as:
\begin{equation}
 \begin{split}
  &\ddot{x}_1+\gamma_1 \dot{x}_1 + \omega_1^2 x_1 + v_{12} x_2 = a_1 e ^{-i\omega t},\\
  &\ddot{x}_2+\gamma_2 \dot{x}_2 + \omega_2^2 x_2 + v_{12} x_1 = 0,
 \end{split}
 \label{eq:motion_eq}
\end{equation}
where $v_{12}$ describes the coupling of the oscillators, $a_1$ is the amplitude of the external force, $\omega$ is the frequency of the force, $\gamma_1, \gamma_2$ are frictional parameters of the oscillators. The steady-state solutions for the displacement of the oscillators are written in form : $x_{1,2}=c_{1,2} e ^{-i\omega t}$, where amplitudes are~\footnote{Note that~\cite{Joe2006} use $x_{1,2}=c_{1,2} e ^{i\omega t}$, so our solution is complex conjugate of the one in~\cite{Joe2006}}~\cite{Joe2006}:
\begin{equation}
    \begin{split}
        &c_1=\frac{(\omega_2^2-\omega^2-i \gamma_2 \omega)}{(\omega_1^2-\omega^2-i \gamma_1 \omega)(\omega_2^2-\omega^2-i \gamma_2 \omega)-v^2_{12}} a_1,\\
        &c_2=-\frac{v_{12}}{(\omega_1^2-\omega^2-i \gamma_1 \omega)(\omega_2^2-\omega^2-i \gamma_2 \omega)-v^2_{12}} a_1.
    \end{split}
    \label{cc}
\end{equation}

\begin{figure}
    \centering
    \includegraphics[width=\linewidth]{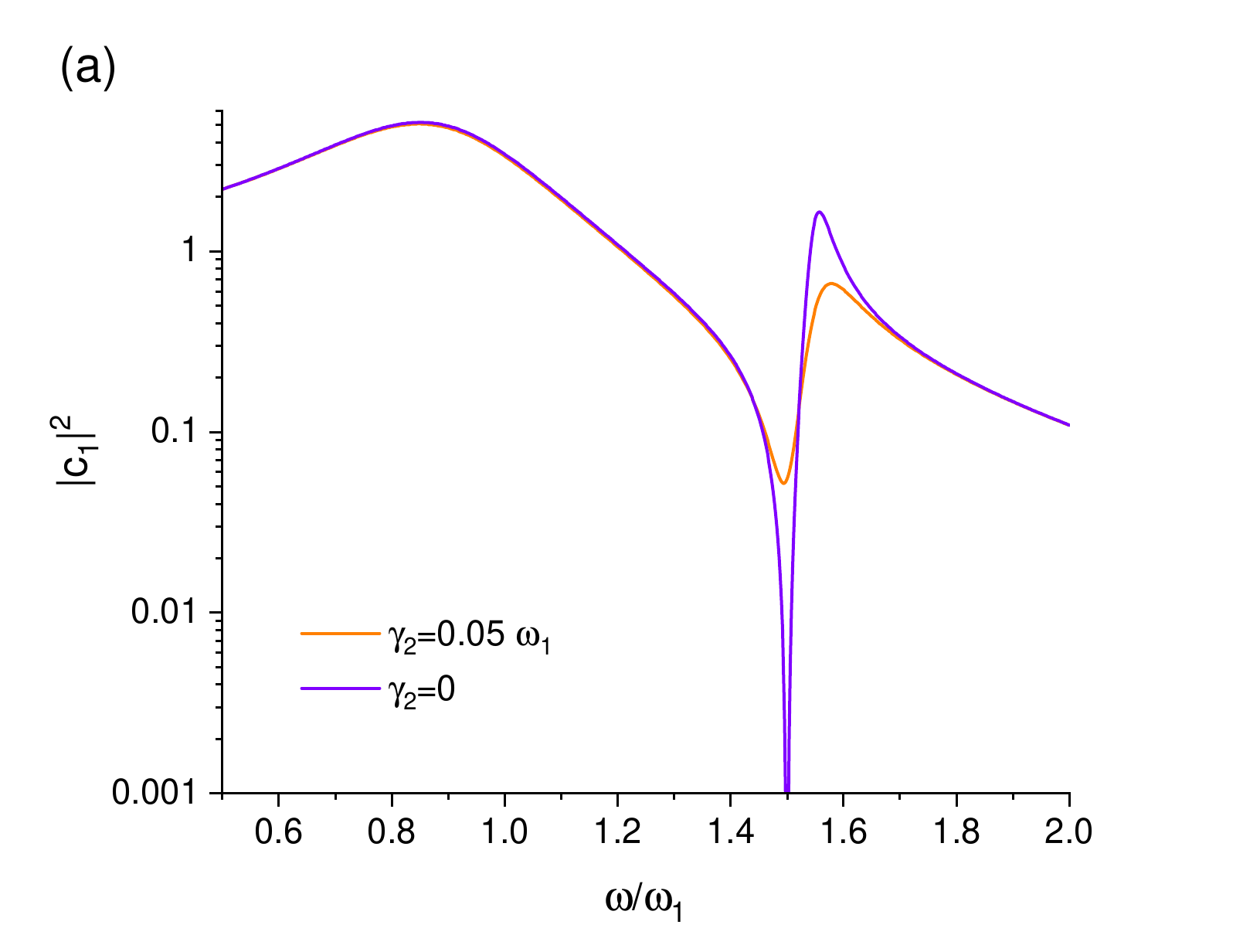}
    \includegraphics[width=\linewidth]{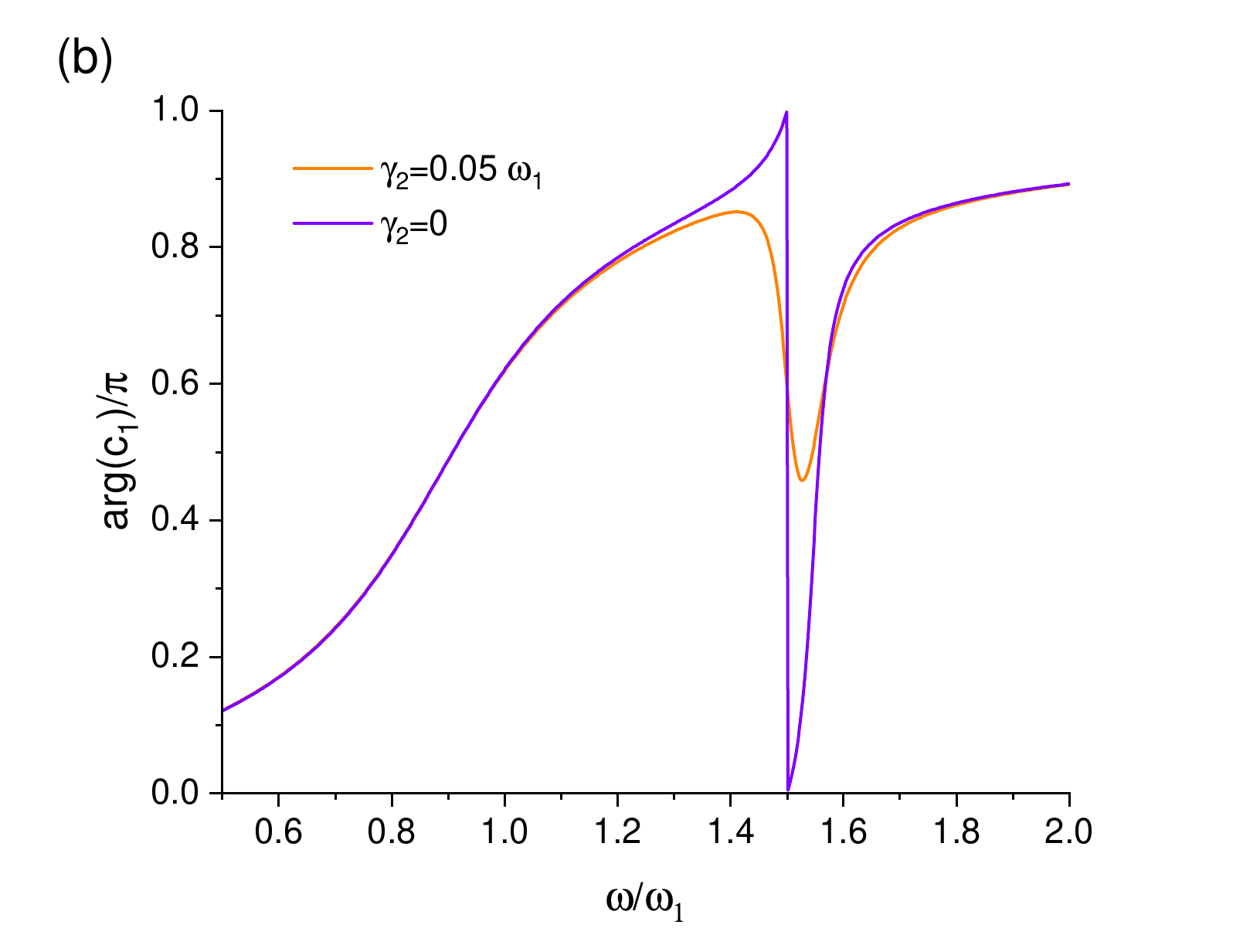}
    \caption{The amplitude (a)  and the phase (b) of the first oscillator motion as a function of the external force frequency under zero and non-zero frictional parameter of the second oscillator. The parameters of the system are: $\omega_2=1.5 \omega_1$, $\gamma_1=0.5 \omega_1$, $v_{12}=0.5 \omega_1$, $\gamma_2$ is shown in the graph.}
    \label{fig:c1}
\end{figure}

The system of the two coupled oscillators has two eigen-frequencies, so the amplitude of the first oscillator's motion as a function the frequency of the external force has two resonant peaks: a symmetric and an asymmetric one. Fig.~\ref{fig:c1} shows the amplitude and phase of the first oscillator's motion as a function of the external force frequency. 

Further we find the resonant factor $F_{cl}(\omega)$ as:
\begin{equation}
    F_{cl} (\omega) =c_1 (\omega)/c_1 (\omega) |_{v_{12}=0}
    \label{F_cl}
\end{equation}
where $c_1 (\omega) |_{v_{12}=0}$ is the amplitude of the first oscillator in the absence of the coupling: 
\begin{equation}
  c_1 (\omega) |_{v_{12}=0}=\frac{a_1}{(\omega_1^2-\omega^2-i \gamma_1 \omega)}.
  \label{c1}
\end{equation}

Let us consider the resonant line near the natural frequency of the second oscillator: $|\omega-\omega_2| \ll |\omega_1-\omega_2|$, so in Eq.~(\ref{cc})-(\ref{c1}) we have: 
\begin{equation}
    \begin{split}
        &\omega_1^2 - \omega^2 \approx \omega_1^2 - \omega_2^2, \\
        &\omega_2^2 - \omega^2 \approx 2 \omega_2 (\omega_2 - \omega), \\
        & \gamma_{1,2} \omega \approx \gamma_{1,2} \omega_2 .\\
     \end{split}
    \label{approx}
\end{equation}
Within these approximations Eq.~(\ref{F_cl}) is equivalent to Eq.~(\ref{eq:fano_line}) with:
\begin{equation}
    \begin{split}
        & Q=\frac{q-i}{1+\gamma_2/\gamma}, \\
        & q=\frac{\omega_2^2-\omega_1^2}{\gamma_1 \omega_2}, \\
        &\gamma=\frac{\gamma_1 \nu_{12}^2}{(\omega_2^2-\omega_1^2)^2+\gamma_1^2 \omega_2^2}, \\
        & \Omega=\omega_2+q \gamma/2 , \\
        & \Gamma=\gamma+\gamma_2 \\
    \end{split}
    \label{parameters_cl}
\end{equation}

Thus, for $\gamma_2=0$ the line has Fano profile with real asymmetry parameter $q$ and for $\gamma_2 \ne 0$ it is characterized with complex asymmetry parameter. For low $\gamma_2$, namely, for $\gamma_2 \ll \gamma$ the asymmetry parameter weakly depends on $\gamma_2$. For higher $\gamma_2$ the parameter $|Q|^2$ decreases with $\gamma_2$ as $1/\gamma_2^{2}$. There is a clear similarity of this behavior and that presented in Fig.~\ref{fig:abs_q}. Namely, $|Q|^2$ decreases with the friction increase in the classical system and with the depletion increase in the quantum one. Thus one can speculate that friction in the coupled oscillators is a classical analogy of the excited state depletion in the quantum system.

\section*{Conclusions}
    In this paper we study generation of coherent XUV with intense few-cycle laser pulse near the resonance with the transition from the ground to an AI state. We integrate numerically the Schr\"odinger equation for 1D helium atom and find the XUV spectra for different durations of the laser pulse. There is an asymmetric resonant peak in these spectra. Approximating this peak with Eq.~(\ref{eq:fano_line}), we find the complex parameter $Q$. We show that $|Q|^2$ decreases with the pulse duration increase. This decrease corresponds to the suppression of the resonant contribution to the XUV spectrum.  Such suppression was demonstrated in experiments on HHG in helium using short laser pulses.  It can be attributed to the photoionization of the AIS by the laser pulse: this photoionization doesn't take place in very short pulse because the field vanishes directly after rescattering, but it does take place in longer pulse. We consider also a classical system demonstrating Fano resonant peak, namely two coupled oscillators. We show that in the absence of friction in the second oscillator the resonant line is described with a real asymmetry parameter.
    The resonant factor is exactly zero under certain detuning and has a phase jump at this point. In the presence of friction in the second oscillator the resonant factor doesn't vanish at any detuning and there is no phase jump.  In this case the resonant line shape is described using  complex asymmetry parameter. The absolute value of the latter decreases with the frictional parameter. Thus, the friction in the coupled oscillators system can be treated as a classical analogy of the extra depopulation of the excited state in the quantum system demonstrating the Fano resonance.

\section*{Acknowledgment}
This study was funded by RSF through Grant No. 24-12-00461. 

\bibliography{lit} 

\end{document}